\documentclass[twocolumn,trackchanges]{aastex631}

\newcommand\msun{M_\odot}

\usepackage{color}

\shorttitle{Mock lobular clusters tidal tails}
\shortauthors{Andr\'es E. Piatti}

\begin{document}

\title{Revisiting physical properties of mock globular cluster tidal tails}

\author[0000-0002-8679-0589]{Andr\'es E. Piatti}
\affiliation{Instituto Interdisciplinario de Ciencias B\'asicas (ICB), CONICET-UNCuyo, Padre J. Contreras 1300, M5502JMA, Mendoza, Argentina}
\affiliation{Consejo Nacional de Investigaciones Cient\'{\i}ficas y T\'ecnicas (CONICET), Godoy Cruz 2290, C1425FQB,  Buenos Aires, Argentina}

\correspondingauthor{Andr\'es E. Piatti}
\email{e-mail: andres.piatti@fcen.uncu.edu.ar}

\begin{abstract}
In this work we present results of the first in-depth analysis of 
extra-tidal mock stars of Milky Way globular clusters recently generated by \citet{grondinetal2024}.
Particularly, we selected
a sample of globular clusters with a general consensus of being formed in the bulge or 
in the disk of the Milky Way. From the catalog we estimated the width and the dispersion 
in the z-component of the angular momentum and in the line-of-sight and tangential 
velocities of their tidal tails, and compared the results with those predicted
by cosmological simulations of \citet{malhanetal2021} and observations.
We found that the resulting values of these four quantities are not in agreement
with an in-situ formation of the associated globular clusters.
On average, the resulting widths agree with an in-situ origin, while
the dispersion in the z-component of the angular momentum, and the dispersion in the 
line-of-sight and in the tangential velocities fail in matching this formation scenario.
The four quantities derived for globular clusters
formed in the bulge or in the disk show similar correlations with the stream length,
namely: the width and the dispersion in the z-component of the angular 
momentum increase with the stream length, while the bulk of dispersion values in the line-of-sight and 
in the tangential velocities is around 12 km s$^{\rm -1}$ along the mock stream.
\end{abstract}


\section{Introduction}

It is known that globular clusters lose stars by tidal stripping during their lifetime, 
although only a percentage of them do already have detected tidal tails \citep{zhangetal2022}. 
Some tidal tails have been targeted by detailed studies which 
showed their complex physical structures \citep{malhanetal2019,deboeretal2020,bonacaetal2020,grillmair2022}. Such
tidal structures are the results of the dynamical evolution of globular clusters
in the host galaxy gravitational field. Recently, \citet{grondinetal2024} 
generated a catalog of mock extra-tidal stars for 159 Milky Way globular clusters, 
providing sky position, kinematics and stellar properties for all simulated stars. 
These extra-tidal mock stars were generated from strong three-body encounters in the core 
of the globular clusters, which produce significantly higher ejection velocities (and dispersions)
than the combination of weak two-body relaxation and tidal stripping that produce most stars
in a tidal tail/stellar stream \citep[e.g.][]{ss72,bt2008,weatherfordetal2023}. 
Since \citet{grondinetal2024}'s mock stars are not generated from accreted halos,
their globular cluster formation reproduces the in-situ scenario. They 
found that mock extra-tidal stars show very good agreement with tidal tail stars of the
globular cluster Pal~13, used as a case study, so that the catalog has been made publicly 
available to help identifying tidal tail stars whenever shallow observational data are available 
or they do not exist.

\begin{deluxetable*}{lccccccccccc}
\tablecaption{Length, width and dispersion in $L_z$, $V_{los}$, and $V_{Tan}$ of tidal tails of
globular clusters formed in the bulge/disk of the Milky Way.}
\label{tab1}
\tablewidth{0pt}
\tablehead{ \colhead{Cluster}  & \colhead{origin} & \colhead{. } & \colhead{. } & \colhead{10$\%$ sample}
&  \colhead{. } & \colhead{. } & \colhead{. } & \colhead{. } & \colhead{50$\%$ sample}
&  \colhead{. } & \colhead{. } \\\hline
 \colhead{}   &  \colhead{}  & \colhead{length} & \colhead{$w$} & \colhead{$\sigma$$_{L_z}$} & 
 \colhead{$\sigma$$_{V_{los}}$} & \colhead{$\sigma$$_{V_{Tan}}$} &
\colhead{length} & \colhead{$w$} & \colhead{$\sigma$$_{L_z}$} & \colhead{$\sigma$$_{V_{los}}$} 
& \colhead{$\sigma$$_{V_{Tan}}$} \\
 \colhead{}  &  \colhead{}  & \colhead{(kpc)}  & \colhead{(pc)} & \colhead{(km s$^{\rm -1}$ kpc)} 
 & \colhead{(km s$^{\rm -1}$)} & \colhead{(km s$^{\rm -1}$)} &
\colhead{(kpc)}  & \colhead{(pc)} & \colhead{(km s$^{\rm -1}$ kpc)} & \colhead{(km s$^{\rm -1}$)} 
& \colhead{(km s$^{\rm -1}$)}}
\startdata
Liller~1  &   bulge   & 2.7  & 200 & 41.4 &  80.0 &  98.4 & 1.1 & 110 & 36.0 &  81.1 & 97.3 \\         
NGC~6093  &   bulge   & 9.8  & 690 & 73.8 &  81.8 &  73.8 & 1.5 & 260 & 55.0 &  54.3 & 41.4 \\
NGC~6144  &   bulge   & 2.9  &  80 & 22.3 &  15.0 &  20.5 & 1.0 &  50 & 17.9 &  12.3 & 14.5 \\
NGC~6171  &   bulge   & 2.0  & 150 & 35.3 &  23.5 &  18.1 & 0.9 &  70 & 31.3 &  14.6 & 13.9 \\
NGC~6266  &   bulge   & 2.5  & 160 & 60.9 &  54.5 &  36.6 & 0.6 &  80 & 56.0 &  47.7 & 31.7  \\
NGC~6293  &   bulge   & 2.0  &  90 & 22.7 &  18.1 &  16.1 & 0.8 &  20 & 12.0 &  10.6 & 10.6  \\  
NGC~6325  &   bulge   & 1.0  &  40 & 11.7 &  18.4 &  13.6 & 0.3 &  10 & 6.2  &  7.9  & 10.4 \\  
NGC~6342  &   bulge   & 1.8  &  60 & 12.9 &  15.0 &  15.9 & 0.4 &  20 & 7.2  &  8.9  & 8.8 \\  
NGC~6380  &   bulge   & 1.0  & 150 & 33.3 &  31.7 &  24.3 & 0.4 &  40 & 34.9 &  16.7 & 16.3 \\  
NGC~6388  &   bulge   & 3.1  & 270 & 132.1&  64.6 &  40.3 & 0.9 & 110 & 142.9&  52.2 & 33.1\\  
NGC~6401  &   bulge   & 1.7  &  60 & 27.4 &  48.6 &  36.6 & 0.4 &  20 & 10.7 &  19.9 & 16.7 \\  
NGC~6440  &   bulge   & 1.3  & 110 & 28.9 &  39.2 &  53.1 & 0.6 &  60 & 26.7 &  27.2 & 43.1 \\  
NGC~6453  &   bulge   & 2.3  & 130 & 35.6 &  29.3 &  22.1 & 0.4 &  30 & 30.1 &  11.6 & 13.8 \\  
NGC~6517  &   bulge   & 1.5  & 150 & 62.7 &  39.2 &  41.9 & 0.9 & 100 & 62.8 &  34.7 & 37.6\\  
NGC~6522  &   bulge   & 1.0  &  40 & 11.2 &  19.5 &  21.2 & 0.4 &  30 & 12.7 &  15.7 & 14.6\\  
NGC~6535  &   bulge   & 1.8  & 130 & 44.0 &  14.8 &  20.5 & 0.4 &  40 & 26.3 &  7.5  & 9.9\\  
NGC~6558  &   bulge   & 1.3  &  50 & 17.4 &  29.3 &  37.0 & 0.4 &  40 & 16.3 &  20.3 & 28.6 \\
NGC~6626  &   bulge   & 5.2  & 680 & 70.4 &  112.7&  59.2 & 0.9 & 140 & 69.3 &  31.5 &  29.3 \\  
NGC~6637  &   bulge   & 2.0  & 100 & 15.6 &  24.0 &  24.7 & 0.7 &  50 & 10.1 &  17.2 & 13.3 \\  
NGC~6652  &   bulge   & 5.8  & 270 & 30.5 &  64.9 &  33.6 & 1.3 & 100 & 22.4 &  36.8 & 14.8 \\  
NGC~6723  &   bulge   & 3.7  &  90 & 19.7 &  41.5 &  35.6 & 0.8 &  40 & 7.6  &  24.5 & 19.4 \\  
Pal~6	  &   bulge   & 4.3  & 150 & 30.0 &  60.1 &  42.0 & 1.0 &  70 & 21.3 &  25.2 & 16.5\\  
Ter~1	  &   bulge   & 2.1  & 200 & 48.0 &  49.1 &  20.0 & 0.5 &  70 & 43.0 &  35.0 & 14.7 \\  
Ter~6     &   bulge   & 2.3  & 760 & 22.7 &  30.3 &  25.4 & 0.4 &  40 & 17.3 &  17.5 & 18.9 \\  
VVV-cl001 &   bulge   & 5.1  & 120 & 60.1 &  66.7 &  43.5 & 0.8 &  90 & 47.6 &  182.6& 70.2 \\  
IC~1276   &   disk    & 13.2 & 240 & 70.0 &  21.5 &  22.7 & 1.5 &  70 & 54.0 &  12.8 & 11.8  \\  
Lyng\aa~7 &   disk    & 3.0  & 180 & 51.6 &  16.4 &  32.7 & 0.7 &  60 & 50.0 &  12.0 & 17.9 \\  
NGC~5927  &   disk    & 12.1 & 400 & 82.5 &  24.9 &  26.4 & 1.2 & 120 & 84.7 &  16.2 & 20.6 \\  
NGC~6218  &   disk    & 3.6  & 220 & 52.4 &  25.6 &  21.6 & 1.8 & 120 & 48.2 &  17.1 & 18.7 \\  
NGC~6352  &   disk    & 2.6  & 120 & 34.2 &  14.5 &  15.3 & 1.5 &  80 & 30.9 &  11.7 & 12.5 \\  
NGC~6362  &   disk    & 3.2  & 160 & 43.6 &  13.0 &  16.6 & 0.8 &  60 & 50.6 &  10.7 & 12.6 \\  
NGC~6366  &   disk    & 3.4  & 170 & 48.2 &  21.1 &  18.3 & 0.7 &  50 & 50.5 &  10.9 & 12.3 \\  
NGC~6441  &   disk    & 5.2  & 890 & 185.6&  74.3 &  44.3 & 2.0 & 340 &174.5 &  49.6 & 37.1 \\  
NGC~6496  &   disk    & 3.8  & 140 & 41.6 &  16.5 &  14.3 & 1.0 &  80 & 31.2 &  12.7 & 11.7 \\  
NGC~6838  &   disk    & 6.6  & 200 & 66.9 &  14.4 &  16.4 & 0.8 &  70 & 44.6 &   7.0 &  8.3 \\  
Pal~10    &   disk    & 27.2 & 830 & 155.0&  35.5 &  24.8 & 4.2 & 300 & 144.6&  18.4 & 17.0 \\  
Pal~11    &   disk    & 4.4  & 140 & 46.9 &  9.3  &  7.2  & 1.1 & 70  & 41.8 &   6.7 &  5.5\\
\enddata
\end{deluxetable*}

In order to assess at what extend these mock tidal tails represent those real ones, and 
therefore are suitable for further tidal tail investigations, we analyzed some of their 
physical properties to the light of the expected values for tidal tails primarily made up 
of lower-speed escapers ejected by two-body relaxation combined with tidal stripping
as found by \citet{malhanetal2021} and \citet{malhanetal2022}. 
These authors proposed that some morphological and dynamical 
properties of globular cluster tidal tails tell us about the origin of Milky Way globular 
clusters, namely, whether they were accreted or formed in-situ. They showed that the width 
of tidal tails, and their dispersion in the $z$-component of the angular momentum, and in the 
line-of-sight and tangential velocities can help disentangling globular clusters' origin. 
Globular clusters formed in a low-mass galaxy halo with cored or cuspy central density profiles 
of dark matter that later merged with the Milky Way develop tidal tails with different mean 
values of the mentioned properties. On average, globular clusters from cuspy profiles have 
tidal tails with the above four properties being three times larger than those of clusters 
formed in cored dark matter profiles. Globular clusters formed in-situ have mean values of 
these quantities nearly ten times smaller than those for globular clusters accreted inside 
cored subhalos. 

The above physical properties are lacking for most of the globular clusters 
with detected tidal tails. As far as we area aware, only M~5 and NGC~288 have recently been targeted 
by \citet{piatti2023b} and \citet{grillmair2025}, respectively, from the selection of  highest ranked 
tidal tail member candidates. \citet{piatti2023b} found from the measurement of the dispersion of
the tangential velocity that M\,5 was accreted from a cuspy $\sim$10$^9$ $\msun$ dark matter 
subhalo, in very good agreement with the overall consensus of being associated to the Helmi
stream \citet[][and reference therein]{callinghametal2022}. Similarly, \citet{grillmair2025} obtained
a tangential velocity dispersion of stream candidates mostly consistent with having been stripped in a 
parent galaxy that had a large, cored dark matter halo. Indeed, NGC~288 is believed to have 
been brought into the Galactic halo during the Gaia-Enceladus-Sausage accretion event
\citep{belokurovetal2018,helmietal2018}. In this context, \citet{grondinetal2024}'s catalog opens 
the possibility to explote whether their mock extra-tidal stars are representative of tidal tails of
globular clusters formed in-situ.  In Section~2, we describe the data handling and
analysis, while in Section~3 we discuss the results. Section~4 summarizes
the main conclusions of this work.

\section{Data analysis}

The globular cluster extra-tidal star catalog generated by \citet{grondinetal2024}
contains 50000 mock stars per globular cluster for a total of 159 globular clusters 
included in \citet{bh2018}. The stars were simulated with the three-body particle spray 
code \texttt{Corespray} 
\citep{grondinetal2023}, with their orbits integrated in the MWPot2014 model available 
in \texttt{galpy} \citep{bovy2015}. From the catalog, we retrieved Right Ascension (R.A.) 
and Declination (Dec.), proper motions along these celestial coordinates 
($\mu_{\alpha}^*,\mu_{\delta}$), line-of-sight velocity ($V_{los}$), heliocentric distance 
($d$), galactocentric distance ($R_{GC}$), 3D galactic coordinates ($X,Y,Z$), and 3D 
action components ($L_z$,$J_R$,$J_z$) for a sample of globular clusters with
a general consensus of being born in the bulge or in the disk of the Milky Way
\citep[see in-situ origin criteria in][and references therein]{callinghametal2022}. We constrained the
globular cluster sample to those with mock tidal tails longer than 1 kpc in any
galactic direction, in order to secure a robust statistical analysis. With the aim
of selecting the most suitable globular clusters, we visually inspected their mock
tidal tails in the ($X,Y,Z$) galactic coordinate space and discarded those with
extra-tidal features extending up to 1 kpc from their respective globular clusters' 
centres. Table~\ref{tab1} lists the resulting globular cluster sample.

\begin{figure*}
\includegraphics[width=\columnwidth]{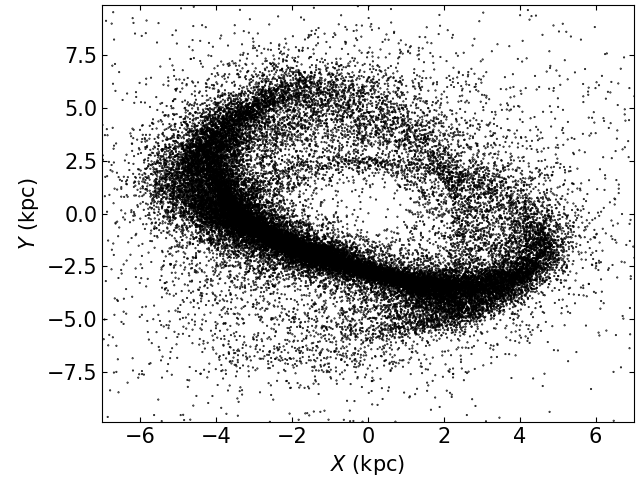}
\includegraphics[width=\columnwidth]{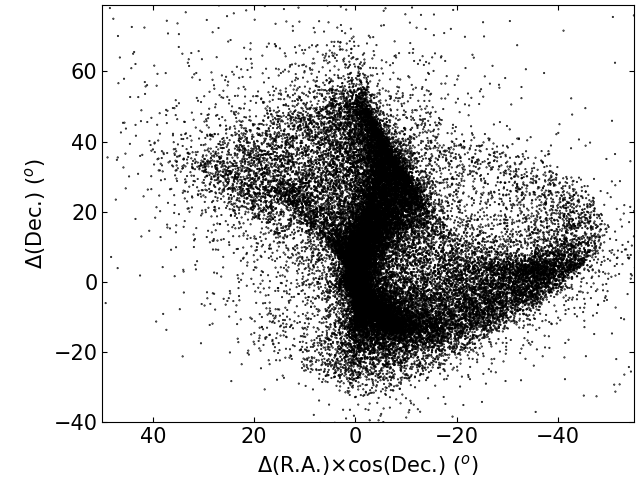}
\caption{Distribution of tidal tail mock stars of NGC~6496 in the Galactic ($X,Y$) plane
(left panel) and in the projected celestial (R.A.,Dec) coordinate system (right panel).}
\label{fig1}
\end{figure*}

One advantage of the \citet{grondinetal2024}'s catalog is that it provides 
Galactic coordinates for each star, so that we can deal with true physical 
dimensions of the tidal tails instead of great circle projections, which 
do not perform satisfactorily for highly radial tidal tails. Because angular
positions of tidal tail stars have mostly been available, the vast majority of observation-based 
tidal tail studies have traced their properties along these angular directions in the sky (e.g.
(R.A., Dec.), ($l$,b) or ($\phi$$_1$,$\phi$$_2$)). Sometimes, 
the mean heliocentric distances of the respective globular clusters have also been
adopted for all tidal tail members \citep[see, e.g.,][and references therein]{ferroneetal2023,mateu2023}.
However, we aim at computing the widths, the dispersion in the $z$-component of the angular momentum, 
and in the line-of-sight and tangential velocities of tidal tails along them, so physical
distances along the tidal tail from the associated globular clusters result more
appropriate. Employing angular distances can result in larger values of the
analysed quantities because of projection effects. Figure~\ref{fig1} depicts the spatial distribution of 
the mock extra-tidal stars of NGC~6496 in the ($X,Y$) Galactic planes (physical spatial
distribution) compared to that in the projected celestial (R.A., Dec.) coordinate system. 
As can be seen,  projection effects are clearly visible in the celestial plane.

With the aim of tracing the properties of interest along the directions of the tidal tails 
of a globular cluster, we  first built a stellar density map in the ($X,Y,Z$) space, and 
superposed on it ten 3D density level contours. The tidal tails of each globular cluster contain 
50000 mock stars, so that the sole number of mock stars located in a region is not indicative 
of the real presence of them. Instead, 
density levels illustrate more properly, statistically speaking, the tidal tail characteristics. 
The highest density 
level is at the globular cluster position. We then centered the coordinate system on the
the globular cluster and performed two perpendicular rotations
in order to have the tidal tails mainly oriented along one of the three perpendicular axes.
Firstly, we rotated the galactic ($X,Y,Z$) system around the $Y$ axis 
to the ($X',Y,\phi_3$) system, as follows:\\

$X' = X cos(\theta) + Z sin(\theta)$\\

$Y$ = Y\\

$\phi_3 = -X sin(\theta) + Z cos(\theta)$,\\

\noindent where $\theta$ is the rotating angle to have the tidal tail in the ($X,Z$) plane
aligned along the $X'$ direction. Then, we rotated the ($X',Y,\phi_3$) system around
the $\phi_3$ axis to the ($\phi_1,\phi_2,\phi_3$) system, as follows:\\

$\phi_1 =  X' cos(\psi) - Y sin(\psi)$\\

$\phi_2 = X' sin(\psi) + Y cos(\psi)$\\

\begin{figure}
\includegraphics[width=\columnwidth]{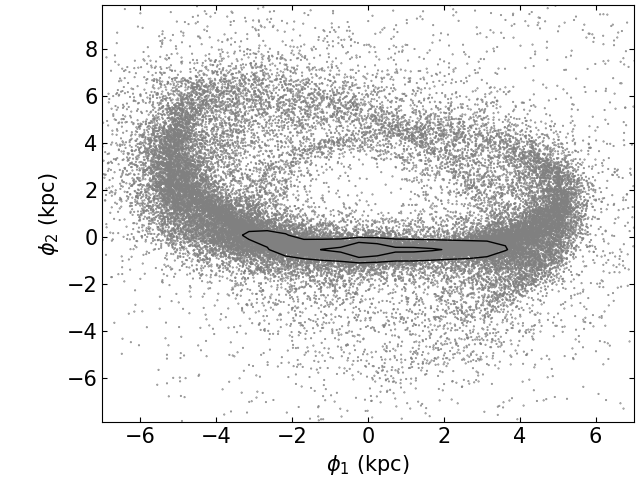}
\caption{Same as left panel of  Figure~\ref{fig1}, once the ($X,Y,Z$) coordinate system
is rotated to the ($\phi$$_1$,$\phi$$_2$,$\phi$$_3$) one. The smaller and larger contours 
correspond to the 50$\%$ and 10$\%$ of the highest stellar density level.}
\label{fig2}
\end{figure}

$\phi_3 = \phi_3$\\,

\noindent where $\psi$ is the rotating angle to have the tidal tail in the ($X',Y$) plane
aligned along the $\phi_1$ direction. The appropriate rotation angles $\theta$ and $\psi$
were obtained by visually inspecting the orientation of the
stellar density contours in the rotated 3D coordinate system. We called the rotated framework 
($\phi$$_1$,$\phi$$_2$,$\phi$$_3$), where $\phi$$_1$ is along the tidal tails, $\phi$$_2$
is perpendicular to $\phi$$_1$ and contained in the tidal tails plane, and $\phi$$_3$ is
perpendicular to $\phi$$_1$ and $\phi$$_2$. Figure~\ref{fig2} illustrates the tidal tails
of NGC~6496 in the ($\phi$$_1$,$\phi$$_2$) plane, with the stellar density levels
corresponding to the 50$\%$ and 10$\%$ of the highest density level superimposed with
a small and a large contour line, respectively. These stars represent the main  
substructures of the tidal tails. We note that tidal tails' stars located
outside the large contour were not used to compute the width, the dispersion in the $z$-component 
of the angular momentum, and in the line-of-sight and tangential velocities of tidal tails.

\begin{figure}
\includegraphics[width=\columnwidth]{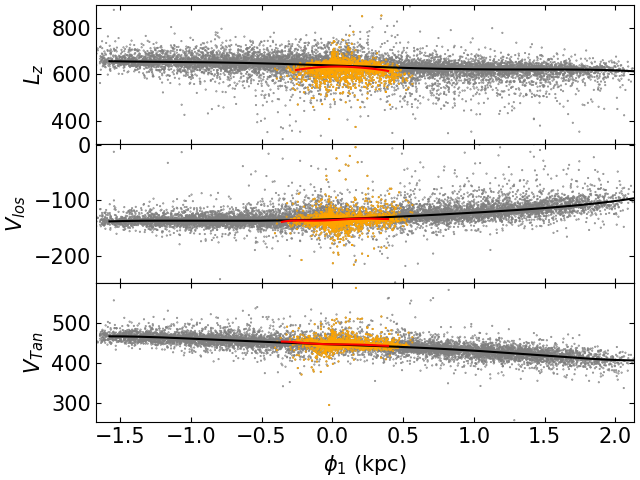}
\caption{Variation of $L_z$, $V_{los}$, and $V_{Tan}$ along the tidal tails of NGC~6496,
measured from the cluster's centre. Grey and orange points represent stars contained
within the stellar density contours corresponding to the 10$\%$ and 50$\%$
of the highest density level, respectively.  Black and red solid lines represent 
the best-fitted polynomials to grey and orange points, respectively.}
\label{fig3}
\end{figure}

The next step consisted in plotting the $z$-component of the angular momentum, and
the line-of-sight and tangential velocities as a function of $\phi$$_1$ for
all the stars located inside the stellar density volume corresponding to the 10$\%$ and 50$\%$
of the highest density level. The tangential 
velocities were computed as $V_{Tan}$ = $k \times d \times \mu$; where $k$ = 4.7405 km 
s$^{\rm -1}$ kpc$^{\rm -1}$ (mas/yr)$^{\rm -1}$, and $\mu = \sqrt{{\mu_{\alpha}^*}^2 + 
\mu_{\delta}^2}$. Figure~\ref{fig3} illustrates the resulting spatial distributions 
of $L_z$, $V_{los}$, and $V_{Tan}$ along the tidal tails direction ($\phi_1$) of NGC~6496 for the
10$\%$ and 50$\%$ samples, represented with grey and orange points, respectively. 
We then fitted the observed distributions with polynomials of up to 5th order (see black 
and red curves in Figure~\ref{fig3} for 10$\%$ and 50$\%$ samples, respectively), which represent 
the mean behaviour of the regarded physical properties along the tails. Finally, we computed 
the residuals (mock individual value - mean fitted value for the respective $\phi_1$)
and the respective standard dispersion, namely: $\sigma$$_{L_z}$, 
$\sigma$$_{V_{los}}$, and $\sigma$$_{V_{Tan}}$, which are quantities proposed by 
\citet{malhanetal2021} and \citet{malhanetal2022} to disentangling the origin of Milky Way 
globular clusters as formed in dwarf galaxies with central cored or cuspy dark matter profiles 
or formed in-situ. The fourth proposed property, the tidal tails width ($w$), was computed from 
the residuals of the spatial distribution of stars in the ($\phi$$_2$,$\phi$$_3$) plane.
Table~\ref{tab1} lists the resulting values for the selected globular clusters with origin
in the bulge or in the disk of the Milky Way.

\section{Discussion}

As mentioned in Section~1, detection of tidal tails of globular clusters has been 
constrained by observational data depth and sky spatial coverage. \citet{zhangetal2022} 
compiled a stringent list of globular clusters with studies of their surrounding fields 
focused on the search of extra-tidal structures. They followed the classification
proposed by \citet{pcb2020} of globular clusters with tidal tails (T), or with
extended envelopes (E), or with no detection of extra-tidal structures (N) to
label NGC~6362 with a T. None of the other studied globular clusters in this work 
(see Table~\ref{tab1}) were included in their compilation. 
According to \citet{grondinetal2024}, the studied globular clusters have 
extra-tidal stars generated by three-body encounters in the clusters' cores 
that are distributed similarly to tidal tail stars.
They can be visualized from the available version of their mock extra-tidal star catalog, accessible 
online at https://zenodo.org/record/8436703 \citep{grondinetal2023catalog}. From their stellar density maps, we found
that the stellar density varies along the tidal tails (along $\phi$$_1$), in 
the sense that the farther the position from the globular cluster, the lower the stellar
density. There is a somehow representative stellar density level --that corresponding
to $\sim$10 $\%$ of the highest density level-- from which
lower values of the stellar density do not follow the coherent stellar 
stream structure.

If we considered the stellar density levels (contours) of the tidal tails,
the 50$\%$ sample would include every star located inside the so-called
half-mass contour. Therefore,
this sample of tidal tail stars is representative of the core features of the
stellar stream. Because the tidal tails extend far beyond the half-mass
contour, we also used stars distributed throughout a
larger extension of the tails to estimate their properties. 
To this respect, we found by inspecting the studied globular cluster tidal tails
that the lowest stellar density level that still
preserves a coherent large-scale tidal tails structure is that corresponding
to the 10$\%$ of the highest density level. For lower stellar density levels
the stellar density maps of the tidal tails only present scattered low
density debris. Because spatially spread and very low stellar density 
sub-structures do not contribute to the main tidal tails features, we
discarded them from our analysis.

\begin{figure*}
\gridline{
\includegraphics[width=\columnwidth]{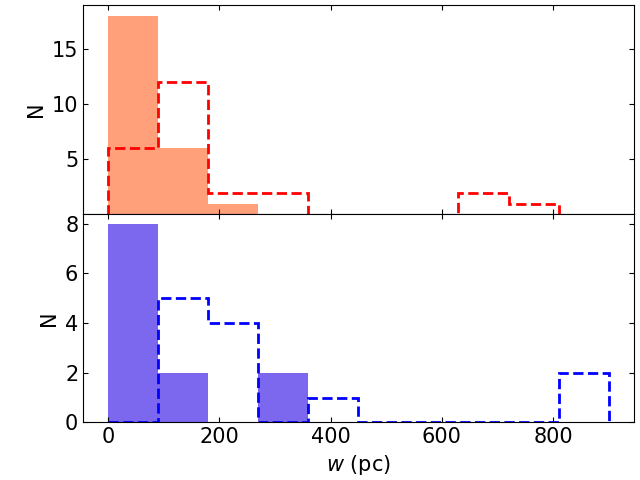}
\includegraphics[width=\columnwidth]{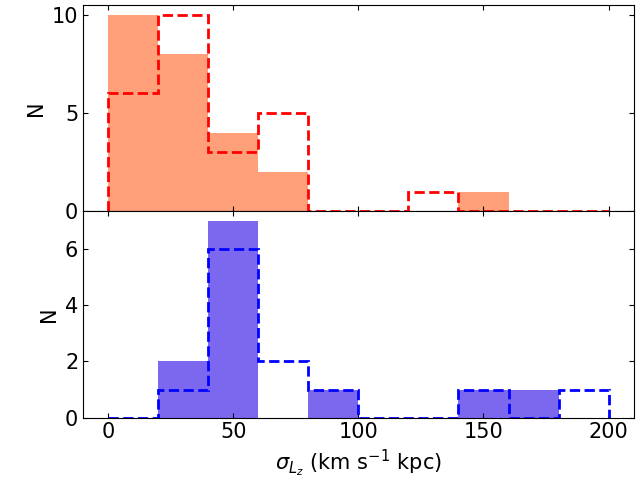}
}
\gridline{
\includegraphics[width=\columnwidth]{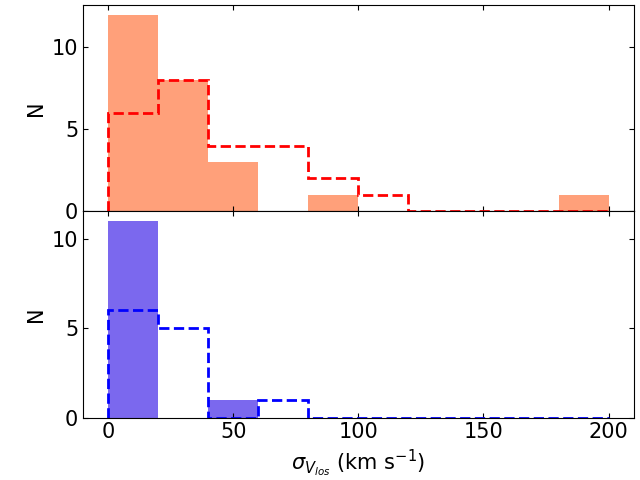}
\includegraphics[width=\columnwidth]{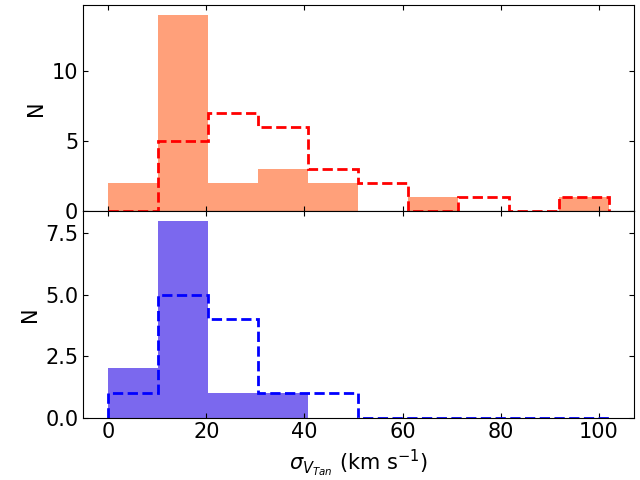}
}

\caption{Distribution of the resulting $w$, $\sigma$$L_z$, $\sigma$$V_{los}$,
and $\sigma$$V_{Tan}$ values for the 10$\%$ and 50$\%$ samples, distinguished
by dashed and solid histograms, respectively. Globular clusters formed in the bulge
or in the disk of the Milky Way are colored red and blue, respectively.}
\label{fig4}
\end{figure*}

Figure~\ref{fig4} shows the resulting distribution of $w$, $\sigma$$L_z$, $\sigma$$V_{los}$,
and $\sigma$$V_{Tan}$ values for the 10$\%$ and 50$\%$ samples, distinguished
by dashed and solid histograms, respectively. Globular clusters formed in the bulge
or in the disk of the Milky Way are colored red and blue, respectively.
We then compared the widths, and the dispersion in the z-component of the angular
momentum, and in the line-of-sight and tangential velocities for the 10$\%$
and 50$\%$ samples. The ratio of the 10$\%$ sample to the 50$\%$
sample for these four properties resulted to be $\sim$ 1, which means that
$w$, $\sigma$$L_z$, $\sigma$$V_{los}$, and $\sigma$$V_{Tan}$ are on average
constant along the considered extensions of the mock tidal tails.
This means that in order to assess at what extend  these mock tidal tails represent those real ones, 
we do not need to choose a stellar
density level cut-off of the tidal tails to be used.

\citet{malhanetal2021} found that the probability distribution functions of their
simulations give values for the tidal tails width of $<$ 100 pc, and within 
 $\sim$ 200-300 pc, 300-400 pc, 700-800 pc, and 1600-2100 pc   ($\pm$3$\sigma$)
for streams formed in-situ, in cored dark matter profiles for masses of 10$^8$$\msun$ 
and 10$^9$$\msun$, and in cuspy dark matter profiles for masses of 10$^8$$\msun$ and
10$^9$$\msun$, respectively. The distribution of the resulting $w$
values (see Figure~\ref{fig4}) shows that only some streams are within the
limits for an in-situ formation. As far as the dispersion in 
the z-component of the angular momentum is considered, 
\citet{malhanetal2021} predicted probability distributions with values $<$ 15 km 
s$^{\rm -1}$ kpc, and within $\sim$30-40 km s$^{\rm -1}$ kpc, 50-90 
km s$^{\rm -1}$ kpc, 100-130 km s$^{\rm -1}$ kpc, and 250-320 km s$^{\rm -1}$ kpc
($\pm$3$\sigma$), respectively, for tidal tails of globular clusters formed in-situ, 
in dwarf galaxies with 10$^8$ and 10$^9$$\msun$ cored profiles of dark matter and in 
galaxies with 10$^8$ and 10$^9$$\msun$ cuspy profiles of dark matter. 
Our values (10$\%$ and 50$\%$ samples) for the studied globular clusters point on 
average to a 10$^8$$\msun$ central cored profile for the dark matter halo of the 
progenitor dwarf galaxy, in disagreement with their generally accepted  in-situ origin
\citep[][and references therein]{callinghametal2022}.

The dispersion in the line-of-sight and tangential velocities 
derived in this work and those obtained by \citet{malhanetal2021} and
\citet{malhanetal2022} brings to the light some differences. They obtained
probability distribution functions for $\sigma$$_{V_{los}}$ and $\sigma$$_{V_{Tan}}$
that nearly overlap; the former being slightly larger. For streams formed in-situ, 
they found values $<$ 1 km s$^{\rm -1}$; for 10$^8$ and 10$^9$$\msun$ cored dark matter 
profiles ($\pm$3$\sigma$) $\sim$1-2 km s$^{\rm -1}$ and $<$5 km s$^{\rm -1}$, and for  
10$^8$ and 10$^9$$\msun$ cuspy dark matter profiles  ($\pm$3$\sigma$) $\sim$7-12 km 
s$^{\rm -1}$. As can be seen, our values are in general inconsistent with any of the above 
values, and therefore are in disagreement with the values for globular clusters
formed in-situ.

\begin{figure*}
\gridline{
\includegraphics[width=\columnwidth]{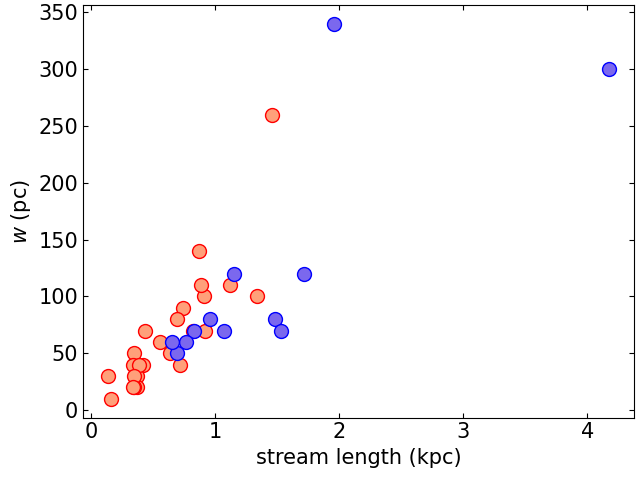}
\includegraphics[width=\columnwidth]{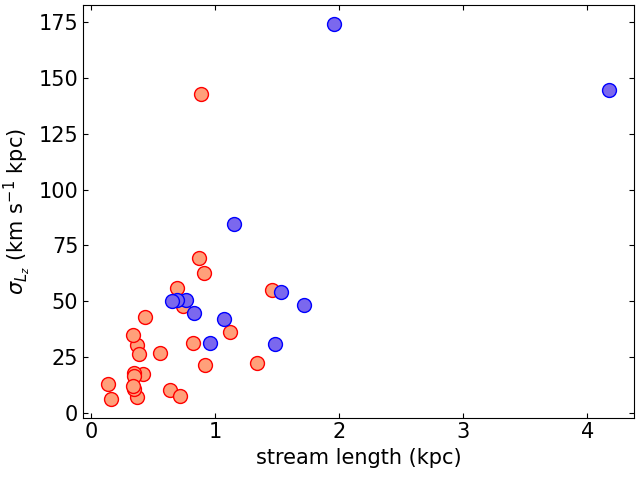}
}
\gridline{
\includegraphics[width=\columnwidth]{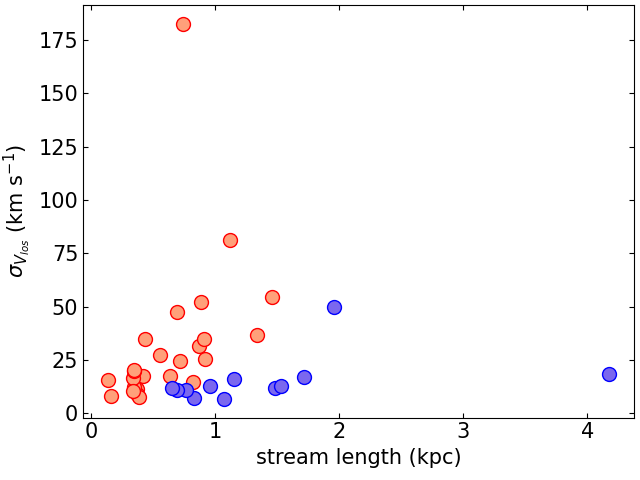}
\includegraphics[width=\columnwidth]{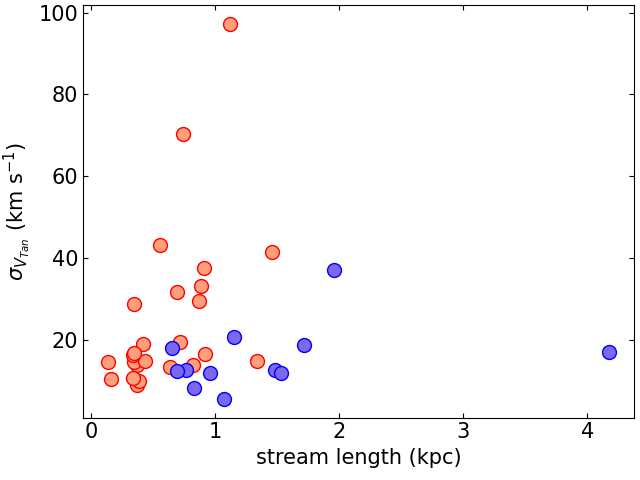}
}
\caption{Relationship of $w$, $\sigma$$L_z$, $\sigma$$V_{los}$,
and $\sigma$$V_{Tan}$ values for the 50$\%$ sample with the stream length, 
measured as the distance between the farthest stars at both side from the globular 
cluster's centre. Globular clusters formed in the bulge or in the disk of
the Milky Way are colored red and blue, respectively.}
\label{fig5}
\end{figure*}

The larger values of velocity 
dispersion derived from the mock extra-tidal stars of \citet{grondinetal2024}
call our attention in the light of recent line-of-sight velocity dispersion
obtained from observational data and numerical simulations. For instance,
\citet{vallurietal2024} recently confirmed $\sigma$$_{V_{los}}$ $<$ 6 km s$^{\rm -1}$
for the GD-1 stream, while \citet{erranietal2022} found $\sigma$$_{V_{los}}$ $\sim$
6 km  s$^{\rm -1}$ for the C-19  stream. 
\citet{piatti2023b} obtained $\sigma$$_{V_{Tan}}$ =
15.65$\pm$0.47 km s$^{\rm -1}$ for M~5, also favoring relatively high values 
for an accreted origin.
However, these result do not reconcile the larger ones for the studied globular clusters 
(Table~\ref{tab1}, 10$\%$ and 50$\%$ samples). Moreover, from 
orbit-averaged Monte Carlo globular cluster simulations and Milky Way-like cold dark 
matter cosmology simulations \citet{weatherfordetal2024} and \citet{ca2024} found 
line-of-sight velocity dispersion of $\sim$ 3-5 km s$^{\rm -1}$ and $<$ 5 km 
s$^{\rm -1}$, respectively, for tidal tails of globular clusters.

We investigated whether there exists any correlation between the widths, and 
the dispersion in the z-component of the angular momentum, and in the line-of-sight 
and tangential velocities with the extension of the tidal tails. In order to
do that, we considered the 50$\%$ sample, although overall behaviors for
the 10$\%$ sample are in very good agreement with the former ones. Figure~\ref{fig5}
depicts these correlations. Firstly, globular clusters formed in the bulge or
in the disk of the Milky Way have their tidal tails extension range overlapped;
those formed in the bulge having also smaller tidal tails. We measured the
extension of the stellar streams by computing the distance between the farthest
stars located at both sides from the globular cluster's centre. Secondly, both
bulge and disk globular clusters show similar trends of the analysed properties 
as a function of the stream length. Particularly, the width of the tidal tails
increases with the stream length, a trend that would seem also be the case for the
dispersion of the z-component of the angular momentum. On the other hand,
the dispersion in the line-of-sight and tangential velocities would not seem 
to increase with the stream length. As can be seen, there is an overall scatter
around $\sigma$$V_{los}$ $\sim$ $\sigma$$V_{Tan}$ $\sim$ 12 km  s$^{\rm -1}$
for any stream length, although much larger values are also obtained for
globular clusters formed in the bulge of the Milky Way.

\section{Conclusions}

The tidal tails of globular clusters contain imprints of their formation and
dynamical evolution. Despite their impact on our understanding of the
formation of the Milky Way, it is still lacking globular clusters with bonafide
tidal tail stars to estimate their physical properties.
Recently, \citet{grondinetal2024} generated a catalog of mock extra-tidal
stars of Milky Way globular clusters, so that we took advantage of them
to assess at what extend they reproduce the values of physical properties 
of tidal tails, such as the tidal tail width, and the dispersion of 
the z-component of the angular momentum, and of the line-of-sight and tangential 
velocities \citep{malhanetal2021,malhanetal2022}. Since \citet{grondinetal2024} 
assumed their globular clusters ejecta evolved fully in-situ, we focused on
globular clusters formed in the Milky Way.

In order to perform such an analysis, we selected a sample of globular clusters 
from the 159 included in the catalog, for which there exist an overall consensus of 
being formed in the bulge or in the disk of the Milky Way. This resulted in a 
self-consistent globular cluster sample suitable to carry out the aforementioned probe. 
We devised two different
mock extra-tidal star samples for each globular cluster aiming at represent
the properties of the tidal tails out to the half-mass density contour
(50$\%$ sample) and out to the outermost coherent substructure (10$\%$ sample), respectively. 
The former tells us about the characteristics of the tidal tail regions closer to its 
accompanying globular cluster, while the latter provides an overall picture of the tidal 
tails. After tracing the ridge line along the physical trailing-leading tails direction,
we computed the four properties mentioned above following the recipes outlined in
\citet{malhanetal2021} and \citet{malhanetal2022} to conclude that:\\

$\bullet$ The width of the mock tidal tails, as well as the dispersion in $L_z$,
$V_{los}$, and $V_{Tan}$ for both 50$\%$ and 10$\%$ samples resulted to be
similar one to each other, respectively.\\

$\bullet$ The values of the width of the tidal tails of only some studied globular clusters
are in agreement with the expected range of values for an in-situ origin in the 
\citet{malhanetal2021,malhanetal2022}'s models.
The dispersion of the z-component of the 
angular momentum point to on average a 10$^8$$\msun$ central cored profile  dark 
matter halo of the progenitor dwarf galaxy, while the dispersion of the line-of-sight and of
the tangential velocities are typically so large as to be inconsistent with any of the models 
from  \citet{malhanetal2021} and \citet{malhanetal2022}, in-situ or otherwise.
Therefore, the mock tidal tails of \citet{grondinetal2024} of globular clusters formed 
in-situ would not seem to be similar to those of \citet{malhanetal2021,malhanetal2022}, 
nor to those studied in some observed globular clusters.
This point may be worth making to the reader, especially as the framing of 
\citet{grondinetal2024}'s and \citet{malhanetal2021,malhanetal2022}'s works
do not make this distinction immediately obvious.\\

$\bullet$ The resulting $L_z$ dispersion values and those for $V_{los}$ and $V_{Tan}$ 
point to either consideration of other alternative formation scenarios for
tidal tails associated to globular clusters formed in-situ to be explored, or a more 
constrained sample of mock extra-tidal stars should be used in order to match the reference
values. The present work does highlight a seeming modeling discrepancy that, while not 
necessarily a problem (due to large differences in the model construction designed to capture 
different regimes of globular cluster ejecta), is still worth 
clarifying to avoid confusion in further studies and avoid application of the relevant models
 in regimes they were not intended for.\\

$\bullet$ Globular clusters formed in the bulge or in the disk of the Milky Way 
span similar stream length ranges and their $w$, $\sigma$$L_z$, $\sigma$$V_{los}$,
and $\sigma$$V_{Tan}$ values show also similar correlations as a function of the
stream length. The
widths and the dispersion in $L_z$ increase with the stream length, while
the bulk of dispersion values in $V_{los}$ and $V_{Tan}$ are $\sim$
12 km s$^{\rm -1}$ along the mock streams.

\begin{acknowledgements}
We thank the referee for the thorough reading of the manuscript and
timely suggestions to improve it.

Data used in this work are available upon request to the author.

\end{acknowledgements}


\end{document}